\documentclass[intlimits,twoside,a4paper]{article}

\usepackage[cp1251]{inputenc}

\usepackage[eqsecnum]{cmpj3}

\issue{2020}{23}{4}{43704}
\doinumber{10.5488/CMP.23.43704}
\title[Influence of $XY$ anisotropy on a magnetoelectric effect in spin-1/2 $XY$ chain in a
transverse magnetic field]
{Influence of $XY$ anisotropy on a magnetoelectric effect in spin-1/2 $XY$ chain in a
transverse magnetic field
}
\author[V. Ohanyan]{V. Ohanyan}
\address{
 Laboratory of Theoretical Physics, and Joint Laboratory of Theoretical Physics --- ICTP Affiliated Centre in Armenia,
          Yerevan State University,
         1 Alex Manoogian Str., 0025 Yerevan, Armenia
}
\date{Received June 15, 2020, in final form July 28, 2020}

\begin{document}

\maketitle

\begin{abstract}
A magnetoelectric effect according to Katsura-Nagaosa-Balatsky mechanism in spin-1/2 $XY$ chain in transverse magnetic field is considered. A spatial orientation of the electric field is chosen to provide an exact solution of the model in terms of free spinless fermions. The simplest model of quantum spin chain demonstrating a magnetoelectric effect, a zero temperature case of the spin-1/2 $XX$ chain in a transverse magnetic field with Katsura-Nagaosa-Balatsky mechanism, is considered. The model has the simplest possible form of the magnetization, polarization and susceptibility functions, depending on electric and magnetic fields in a most simple form. For the case of arbitrary $XY$ anisotropy, a non-monotonous dependence of  magnetization on the $XY$ anisotropy parameter is figured out. This non-uniform behaviour is governed by the critical point which is connected with the possibility to drive the system gapless or gapped by the electric field. Singularities of the magnetoelectric susceptibility at the critical value of system parameters are shown.
\end{abstract}

\section{Introduction}

 Magnetoelectrics are materials having both dielectric polarization and magnetization in a single phase and exhibiting a magnetoelectric effect (MEE), a vast class of phenomena of intercoupling of magnetization and polarization in matter \cite{mee1,mee2,dong15,don19}. These materials are particularly important for their application in spintronic devices \cite{app1, app2}. The MEE is a class of phenomena in solids, which can be detected as magnetic field dependance of dielectric polarization and electric field dependance of magnetization. In most interesting cases of non-trivial MEE, the magnetization (dielectric polarization) can be induced by only applying an electric (magnetic) field. Nowadays, several microscopic mechanisms of the MEE are known \cite{mee1,mee2,dong15,don19}. One of these mechanisms is based on the so-called spin-current model or inverse Dzyaloshinskii-Moriya (DM) model and was proposed in a seminal paper by Katsura, Nagaosa and Balatsky~\cite{KNB1}. The Katsura-Nagaosa-Balatsky (KNB) mechanism establishes a connection between the dielectric polarization of the crystal structure unit consisting of two magnetic ions chemically bonded to one or more $p$-elements and the spin states of the ions \cite{KNB1, KNB2}. The dielectric polarization that induces into the bond between two spins in this model is given by the following expression:
\begin{eqnarray}\label{KNB}
\mathbf{P}_{ij}=\mu \mathbf{e}_{ij}\times\mathbf{S}_i\times\mathbf{S}_j,
\end{eqnarray}
here, $\mathbf{e}_{ij}$ is the unit vector pointing from site $i$ to site $j$, and $\mu$ is a microscopic constant characterizing the quantum chemical features of the bond between two metallic ions and $p$-element(s) \cite{KNB1,KNB2}. $\mathbf{S}_i$ and $\mathbf{S}_j$ are the spin operators of the corresponding ion states. The simplest case of the KNB mechanism is the linear arrangement of magnetic ions (spins), the geometrically linear spin chain. If we suppose the chain to be directed toward the $x$-axis, then the local polarization according to equation (\ref{KNB}) acquires the following components:
\begin{eqnarray}\label{Plin}
&&P_{j,j+1}^x=0,\\
&&P_{j,j+1}^y=\mu\left(S_j^yS_{j+1}^x-S_j^x S_{j+1}^y\right), \nonumber \\
&&P_{j,j+1}^z=\mu\left(S_j^zS_{j+1}^x-S_j^x S_{j+1}^z\right). \nonumber
\end{eqnarray}

To a large family of magnetoelectric materials belong those which feature a one-dimensional arrangement of exchange-interaction paths between Cu$^{2+}$ ions, with ferromagnetic nearest-neighbour ($J_1<0$) and antiferromagnetic next-to-nearest neighbour ($J_2>0$) interactions. The corresponding model of one-dimensional spin-1/2 $J_1-J_2$ spin chain is usually referred to as multiferroic spin chain. A list of magnetic materials successfully described by this model is quite  broad:  LiCuO$_2$ \cite{LiCu2O21,LiCu2O22,LiCu2O23,LiCu2O24}, LiCuVO$_4$ \cite{LiCuVO41,LiCuVO42,LiCuVO43}, CuCl$_2$ \cite{CuCl2}, CuBr$_2$ \cite{CuBr21,CuBr22}, PbCu(SO$_4$)(OH)$_2$ \cite{PbCu1,PbCu2}, CuCrO$_4$ \cite{CuCrO4}, SrCuTe$_2$O$_6$ \cite{SrCuTe2O6} just to mention a few of them.

The recent few years were marked by an interest toward exact and numerical investigation of one-dimensional quantum spin models with KNB mechanism. Exact description (for some models supplemented with numerics) for the MEE is available so far only for simplified models, such as strictly linear integrable $XXZ$ chain \cite{bro13}, the same system but in both longitudinal and transverse fields \cite{thakur18}, spin-1/2 $XYZ$ chain \cite{XYZ}, the spin-1/2 $XY$ chain with three-spin interaction \cite{mench15, sznajd18, sznajd19}, generalized quantum compass model with magnetoelectric coupling \cite{oles}, spin-1/2 Heisenberg-Ising ladder \cite{stre20}. However, exact results are very helpful for understanding the general features of the phenomena. Moreover, some of them can serve as a mean-field approximation for the more realistic models. The latter case is typical of a class of exactly solvable spin chain models, where spins interact with each other via two coplanar components (usually taken as $S^x$ and $S^y$).

  In the present paper we focus on the MEE in a $XY$ chain, which was introduced in seminal paper of Lieb, Schulz and Mattis \cite{LSM}. Since KNB mechanism is essentially affected by the physical form of the lattice, the simplest case corresponds to the linear arrangement of spins, which features the polarization given in equation (\ref{Plin}). Then, seeking for exactly solvable case, we have to chose an electric field to be pointed in $y$-direction. This leads to a model of $XY$ chain with DM terms, where DM-vector is parallel to $z$-axis. This model is well known \cite{kon67, sis74, sis75, perk75,zv90,zv91a,zv91b,zv95,zv05, JWXY1, JWXY2, JWXY3, JWXY4, JWXY5, JWXY6, JWXY7, JWXY8,meh14}, though for the recent half a century quite restricted amount of papers have been devoted to it. The paper is organized as follows: in the second section, the formulation of the model and its exact solution in terms of Jordan-Wigner fermionization is given, in the next section the zero temperature MEE for the simplest model of MEE in quantum spin chains is described, then the finite temperature MEE and the effects of the $XY$ anisotropy $\gamma$ are analyzed. The paper is ended with a conclusion.

\section{The model and its exact solution}
Let us consider spin-1/2 $XY$ chain which has a linear form and spin dependent polarization due to KNB mechanism. Supposing the chain to be collinear with the $x$-axis and the electric field to be pointed in $y$-direction according to equation (\ref{Plin}), we arrive at the following Hamiltonian:
\begin{eqnarray}\label{ham1}
\mathcal{H}=J\sum_{j=1}^N\left\{(1+\gamma)S_j^x S_{j+1}^x+(1-\gamma)S_j^y S_{j+1}^y\right\}+E\sum_{j=1}^N \left(S_j^x S_{j+1}^y-S_j^y S_{j+1}^x\right)-B\sum_{j=1}^N S_j^z\,,
\end{eqnarray}
where $S_j^{\alpha}$ are the spin-1/2 operators at lattice site $j$, $E$ is the magnitude of the electric field written in proper units (with coefficient $\mu$ absorbed in it) and $B$ is an external magnetic field pointing in $z$-direction. Various aspects of this model have been considered in a series of papers during the recent decades \cite{kon67, sis74, sis75, perk75,zv90,zv91a,zv91b,zv95,zv05, JWXY1, JWXY2, JWXY3, JWXY4, JWXY5, JWXY6, JWXY7, JWXY8,meh14}. In the present paper we are interested in the MEE in this model, and particularly, in the effects of $XY$ anisotropy~$\gamma$. The model is exactly solvable within the Jordan-Wigner fermionization. To proceed we should first perform a Jordan-Wigner transformation from spin operators to the creation and annihilation operators of lattice spinless fermions:
\begin{eqnarray}\label{JW}
S_j^-=\re^{\ri\piup\sum_{l=1}^{j-1}c_l^+c_l}c_j, \;\; S_j^+=(S_j^-)^+,\;\;S_j^z=c_j^+c_j-1/2,
\end{eqnarray}
where $S_j^{\pm}=S_j^x \pm \ri S_j^y$. In terms of Fermi operators, the Hamiltonian reads:
\begin{eqnarray}\label{ham2}
\mathcal{H}=\sum_{j=1}^N\left\{\frac{J+\ri E}{2}c_j^+ c_{j+1}-\frac{J-\ri E}{2}c_jc_{j+1}^+\frac{J\gamma}{2}(c_J^+c_{j+1}^+-c_jc_{j+1})-B(c_j^+c_j-1/2)\right\}.
\end{eqnarray}
Here, periodic or anti-periodic boundary conditions are assumed, depending on the number of spinless fermions which is a conserved quantity. For even (odd)  particle number, the anti-periodic (periodic) boundary conditions for Fermi operators are imposed, $c_{j+N}=-c_j$ ($c_{j+N}=c_j$). A further step toward  diagonalization of the Hamiltonian is a Fourier transformation,
\begin{eqnarray}\label{ck}
c_j=\frac{1}{\sqrt N}\sum_k \re^{-\ri jk} c_k, \;\quad c_k=\frac{1}{\sqrt N}\sum_j \re^{\ri jk} c_j,
\end{eqnarray}
here, $k$ takes $N$ values in the first Brillouin zone, $-\piup\leqslant k<\piup$, and is equal to $\frac{2\piup}{N}n$ for periodic boundary conditions or $\frac{2\piup}{N}(n+1/2)$ for the antiperiodic ones. Here, $n=-N/2, -N/2+1, ..., N/2-1$ when $N$ is even and $n=-(N-1)/2 -(N-1)/2+1, ..., (N-1)/2-1$ when $N$ is odd.
Then, the Hamiltonian takes an appropriate matrix-form, which is straightforward for diagonalization:
\begin{eqnarray}\label{ham3}
\mathcal{H}=\frac{1}{2}\sum_{-\piup\leqslant k<\piup} \left(c_k^+, c_{-k}\right)\left( \begin{array}{cc} \varepsilon(k) & -\ri J \gamma\sin k \\
                                                                             \ri J \gamma \sin k & -\varepsilon(-k) \end{array} \right)\left(\begin{array}{c}c_k \\
                                                                                                          c_{-k}^+\end{array}\right),
\end{eqnarray}
where $\varepsilon(k)=J\cos k+E \sin k -B$.  Let us perform Bogoliubov transformation to new Fermi creation and annihilation operators,
\begin{eqnarray}\label{beta}
\left(\begin{array}{c} c_k \\ c_{-k}^+ \end{array}\right)=\left( \begin{array}{cc} \ri u_k & v_k \\ -v_k & -\ri u_k\end{array}\right)\left(\begin{array}{c}\beta_k \\ \beta_{-k}^+\end{array}\right),\;\;\left(\begin{array}{c} \beta_k \\ \beta_{-k}^+ \end{array}\right)=\left( \begin{array}{cc} -\ri u_k & -v_k \\ v_k & \ri u_k\end{array}\right)\left(\begin{array}{c}c_k \\ c_{-k}^+\end{array}\right),
\end{eqnarray}
where $u_k^2+v_k^2=1$. Then, putting
\begin{eqnarray}
&&u_k=\frac{1}{\sqrt 2}\sqrt{1+\frac{J\cos k -B}{\lambda_k}},\; \quad v_k=\mbox{sgn}(J\gamma\sin k)\frac{1}{\sqrt 2}\sqrt{1-\frac{J\cos k -B}{\lambda_k}},\nonumber \\
&&\lambda_k=\sqrt{\left(J\cos k -B\right)^2+J^2\gamma^2\sin^2 k}\,,\nonumber
\end{eqnarray}
we finally obtain the diagonal form of the Hamiltonian expressed in terms of free spinless fermions:
\begin{eqnarray}\label{hamdiag}
&&\mathcal{H}=\sum_{-\piup\leqslant k<\piup}E_{\gamma}(k)\left(\beta_k^+\beta_k-1/2\right), \\
&&E_{\gamma}(k) \nonumber=E\sin k+\mbox{sgn}(J\cos k-B)\lambda_k. \nonumber
\end{eqnarray}
For the isotropic $XX$-chain with DM-terms ($\gamma=0$), one can easily see that the Hamiltonian (\ref{ham3}) is already diagonal in $c_k$ operators:
\begin{eqnarray}\label{Hxx}
&&\mathcal{H}_{XX}=\sum_{-\piup\leqslant k<\piup}E_0(k)\left(c_k^+c_k-1/2\right), \\
&&E_0(k)=\sqrt{J^2+E^2} \cos\left(k-\phi\right)-B,\;\;\phi=\arcsin \frac{E }{\sqrt{J^2+E^2}}. \nonumber
\end{eqnarray}
\section{Zero-temperature properties and MEE}
Let us first describe zero-temperature properties of the spin-1/2 $XX$ chain in
presence of the electric and magnetic fields. The simplest quantum chain model exhibiting MEE via KNB mechanism is the system described by the Hamiltonian (\ref{Hxx}). The free-fermion picture here is quite simple. The DM-term breaks a time-reversal symmetry, $E_0(-k)\neq E_0(k)$, and the two Fermi points are not symmetric with respect to $k=0$. They are given by
\begin{eqnarray}\label{k12}
k_{1,2}=\phi\mp \arccos\frac{B}{B_c}, \;\; B_c=\sqrt{J^2+E^2},
\end{eqnarray}
when $-B_c<B<B_c$. For $B\leqslant -B_c$ and $B\geqslant B_c$ all $N$ free-fermion states in the system are occupied or empty, respectively, the ground state energy per one site. Thus, the ground state energy per one site is given by
\begin{eqnarray}\label{e0}
e_0=
\left\{
\begin{array}{lll}
-\frac{B}{2},  &B\geqslant B_c \\
-\frac{B}{2}+\frac{1}{\piup}\left(B\arccos\frac{B}{B_c}-\sqrt{B_c^2-B^2}\right),  &-B_c\leqslant B \leqslant B_c \\
\frac{B}{2},  &B\leqslant -B_c. \end{array}\right.
\end{eqnarray}
Using standard relations, $m_0=-\frac{\partial e_0}{\partial B}$ and $p_0=-\frac{\partial e_0}{\partial E}$, one can find zero-temperature asymptotic values of  magnetization and polarization:
\begin{eqnarray}\label{m0p0}
m_0=
\left\{
\begin{array}{lll}
\frac{1}{2},  B\geqslant B_c \\
\frac{1}{2}-\frac{1}{\piup}\arccos\frac{B}{B_c}, & -B_c\leqslant B \leqslant B_c \\
-\frac{1}{2}, & B\leqslant -B_c. \end{array}\right., \;\;
p_0=
\left\{
\begin{array}{lll}
0, & B\geqslant B_c \\
\frac{E\sqrt{J^2+E^2-B^2}}{\piup(J^2+E^2)}, & -B_c\leqslant B \leqslant B_c \\
0, & B\leqslant -B_c. \end{array}\right.
\end{eqnarray}
A common feature of the free-fermion models with KNB mechanism is that the dielectric polarization becomes zero for empty as well as for full fermionic filling. Besides the magnetic and dielectric susceptibilities, $\chi=\frac{\partial m_0}{\partial B}$ and $\chi_P=\frac{\partial p_0}{\partial E}$, the magnetoelectric systems have one more important quantity to describe the response, a magnetoelectric or mixed susceptibility, which in general case is defined by the following relation:
\begin{eqnarray}\label{al_def}
\alpha_{ij}=\left(\frac{\partial M_i}{\partial E_j}\right)_{T, \mathbf{B}}=\left(\frac{\partial P_j}{\partial B_i}\right)_{T, \mathbf{E}},
\end{eqnarray}
where $M_i(P_j)$ and $B_i(E_j)$ are components of the magnetization (polarization) vector of the sample and external magnetic (electric) fields, respectively. In our case, all susceptibilities are non-zero only within $ -B_c\leqslant B \leqslant B_c$ and are given by
\begin{eqnarray}
\chi=\frac{1}{\piup\sqrt{B_c^2-B^2}}\,, \quad \chi_P=\frac{J^2(B_c^2-B^2)+E^2 B^2}{\piup B_c^4 \sqrt{B_c^2-B^2}}\,,\quad \alpha=-\frac{EB}{\piup B_c^2\sqrt{B_c^2-B^2}}\,,
\end{eqnarray}
respectively. An important feature of the simplest spin-1/2 $XX$ chain
accounting for MEE  vanishes $\alpha$ when any of the two fields, electric or
magnetic, becomes zero. This is an example of trivial MEE, when magnetic (electric) field affects polarization (magnetization) but cannot induce it unless the other field is non-zero. It can be easily seen that at critical magnetic field, $B=\pm B_c$, all susceptibilities have inverse square-root singularities, which are universal properties of $XY$-type chains. The zero temperature magnetization curve around critical field, $B=\pm B_c$ shows a square-root behavior. Although finite-temperature MEE in $XX$ chain was briefly described in  \cite{bro13}, as a limiting case of $XXZ$ chain, the zero-temperature MEE is also worth studying, since this is the simplest example of the MEE, described by simple analytic expressions.
\begin{figure}[!t]
\centerline{\includegraphics[width=0.93\textwidth]{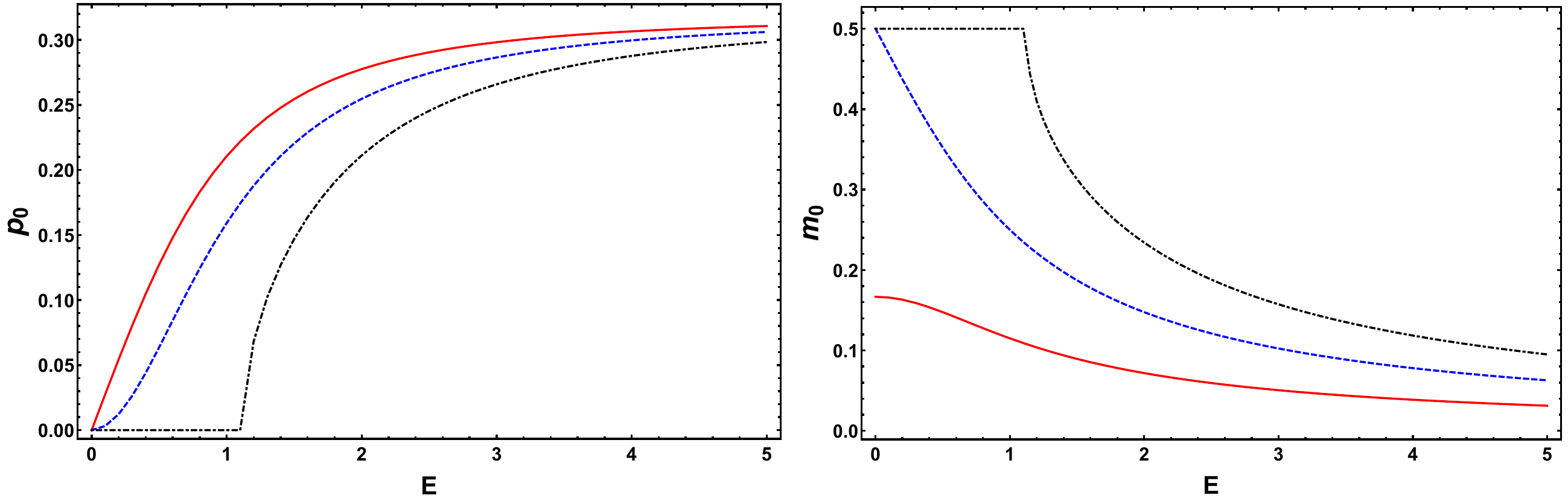}}
\caption{(Colour online) Zero-temperature polarization (left-hand panel) and magnetization (right-hand panel) dependance on electric field for the spin-1/2 $XX$ chain with KNB mechanism. Three regimes of initial polarization $p_0(E)$ are presented: linear ($B<J$), quadratic ($B=J$) and plateau with further square root ($B>J$) at small $E$. The same picture can be seen in the magnetization dependence on the electric field. For both panels $J=1$, red solid line corresponds to $B=0.5$, blue dashed line to $B=1$ and black dot-dashed to $B=1.5$.} \label{fig1}
\end{figure}
In figure \ref{fig1}, zero-temperature polarization and magnetization of the spin-1/2 $XX$ chain with KNB mechanism are presented as functions of electric field. The polarization curves, $p_0(E)$, demonstrate three different regimes of polarization processes close to $E=0$: linear, square-root and plateau with further quadratic behavior. The regime of polarization curve depends on the value of the magnetic field. It is very easy to see from the equation (\ref{m0p0}) that initially polarization curve  has a linear behavior for $B<J$, which becomes quadratic at $B=J$ and then changes to plateau with square-root for $B>J$.  Interestingly, the simplest model of KNB magnetoelectric to a great extent reproduces three of the four qualitative shapes of polarization curves for more complicated spin-1/2 $XXZ$ chain with KNB mechanism \cite{bro13}. The right-hand panel of the figure \ref{fig1} demonstrates magnetization dependence on the electric field for the same three values of magnetic field as in the left-hand panel, $B=0.5$ (red solid line), $B=1$ (blue dashed line) and $B=1.5$ (black dot-dashed line). Obviously, the electric field works against magnetization in the same way as the magnetic field works against  polarization. Zero polarization corresponds to the fully polarized magnetic state given by the plateau at $M=1/2$ at $B>J$ (black dot-dashed line). When $B$ is decreasing, the length of the plateau becomes smaller and reaches zero at critical value $B=J$ (blue dashed line). The red solid line demonstrates a monotonous decrease of magnetization with increasing the electric field for $B<J$.   The plots of zero-temperature magnetization and polarization dependence on a magnetic field are presented in figure \ref{fig2}. Here, the magnetization curves, $m_0(B)$, for different values of the electric field have the same standard form. Moreover, the polarization dependence on the magnetic field is uniform with plateau at $p_0=0$, which corresponds to the fully polarized spin state realized at strong enough magnetic fields.
\vspace{-1mm}
\begin{figure}[!t]
\centerline{\includegraphics[width=0.93\textwidth]{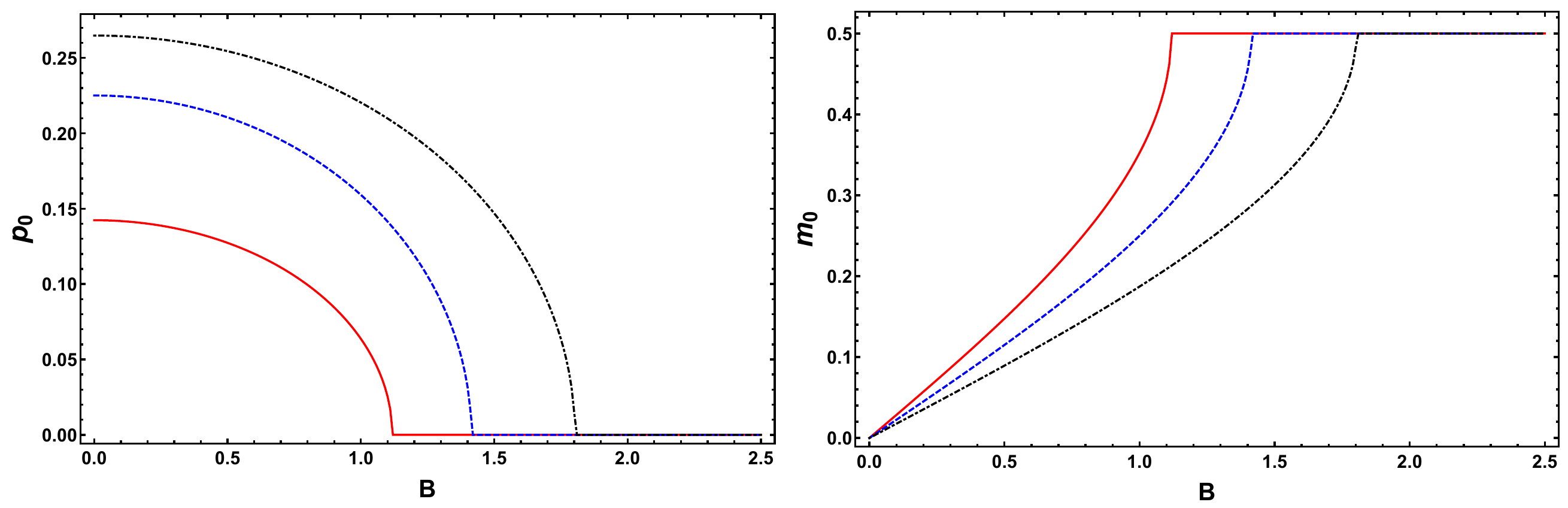}}
\caption{(Colour online) Zero-temperature polarization (left-hand panel) and magnetization (right-hand panel) dependance on magnetic field for the spin-1/2 $XX$ chain with KNB mechanism at $J=1$. For both panels, red solid line corresponds to $E=0.5$, blue dashed line to $E=1$ and black dot-dashed to $E=1.5$.} \label{fig2}
\end{figure}

Zero-temperature description of a more general spin-1/2 $XY$ chain with $\gamma \neq 0$
 is much more complicated \cite{perk75, JWXY1, JWXY2, JWXY3, JWXY4, JWXY5, JWXY6, JWXY7}. In the absence of DM terms, the spectrum is always non-negative, which means that all one-particle free-fermion states are empty. For the spectrum of $XY$ chain with DM terms, given by equation (\ref{hamdiag}), the one-particle excitations are gapless when $E^2\geqslant J^2\gamma^2$ and $B^2\leqslant E^2+J^2(1-\gamma^2)$, or in case of $E^2<J^2\gamma^2$ for $B^2=J^2$. In the latter case, all fermionic one-particle states are still empty, but the spectrum touches zero at a single point. The general property of the MEE in free-fermion models is a vanishing polarization in both cases of fully filled or empty system. Thus, in case of non-zero $\gamma$, polarization is non-zero only in the region of $(E, B)$-plane, given by the conditions, $E^2\geqslant J^2\gamma^2$ and $B^2\leqslant E^2+J^2(1-\gamma^2)$.

\section{Thermodynamics and MEE}
\begin{figure}[!t]
\centerline{\includegraphics[width=0.95\textwidth]{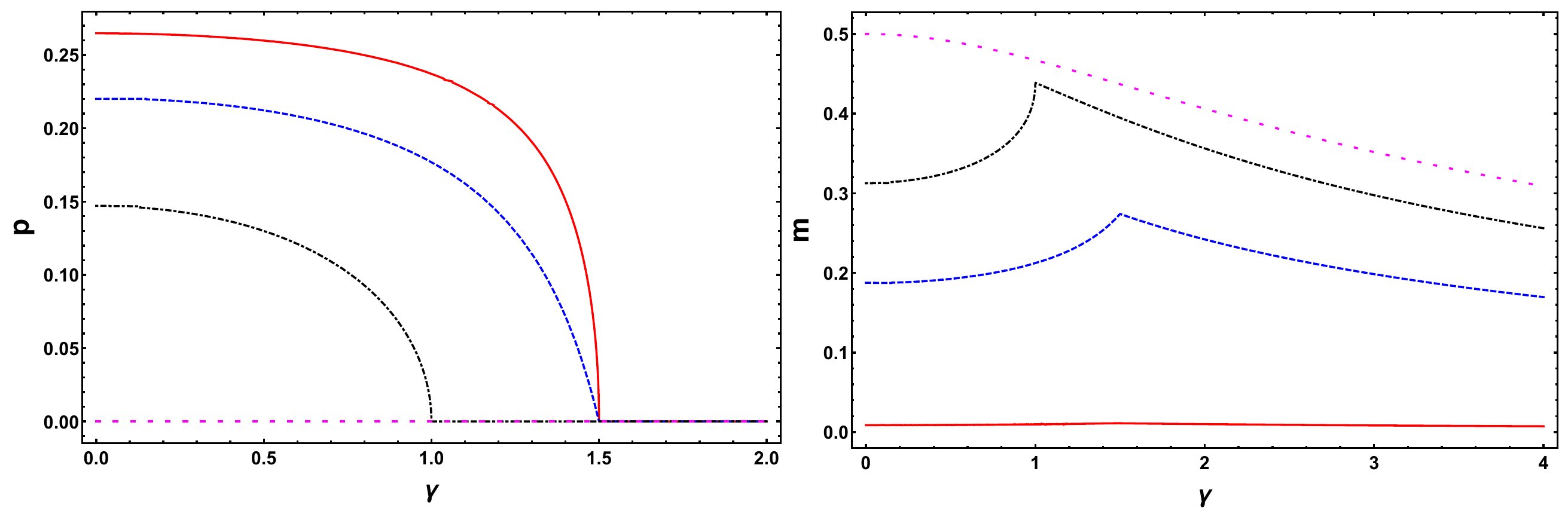}}
\caption{(Colour online) Effect of $XY$ anisotropy parameter $\gamma$ on polarization (left-hand panel) and magnetization (right-hand panel) in appropriate units ($J=1$) and $T=0.001$. Non-monotonous dependence of magnetization as a function of $\gamma$ is clearly seen in the right-hand panel. For both panels $E=1.5$ and $B=0.05$ (red solid line), $B=1$ (blue dashed line) and $B=1.5$ (black dot-dashed line) and $B=2$ (magenta dotted line).} \label{fig3}
\end{figure}

\begin{figure}[!t]
\centerline{\includegraphics[width=0.95\textwidth]{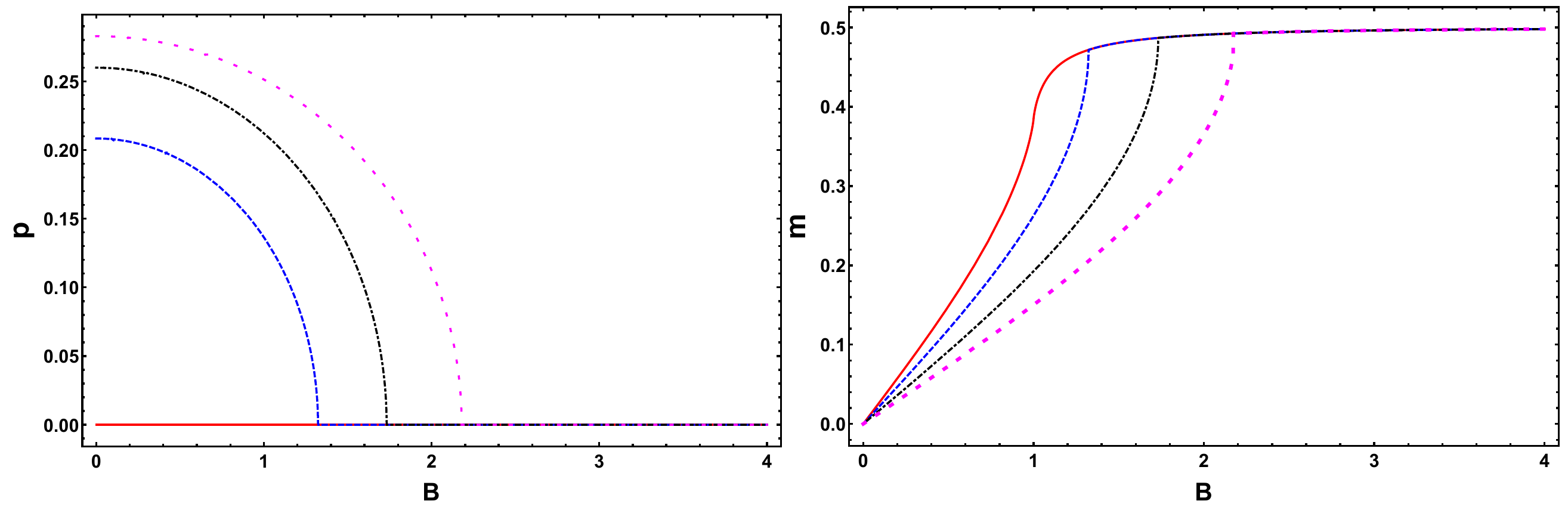}}
\caption{(Colour online) Low-temperature polarization (left-hand panel) and magnetization (right-hand panel) dependence on the magnetic field for the spin-1/2 $XY$ chain with KNB mechanism at $J=1$ in appropriate units. For both panels T=0.0001, $\gamma=0.5$ and $E=0$ (red solid line), $E=1$ (blue dashed line) and $E=1.5$ (black dot-dashed line) and $E=2$ (magenta dotted line).} \label{fig4}
\end{figure}

In the more general case of spin-1/2 XY chain with non-zero $\gamma$,
 it is much  simpler to deal with the thermodynamics of the model than with the zero-temperature expressions, which are quite cumbersome and complicated even for $E=0$, when the system is always gapped or gapless with zero occupation \cite{TM85}. In order to investigate the finite-temperature features of the MEE in the model, we need to start from free energy (per one spin), which is given by the following integral over the first Brillouin zone:
\begin{eqnarray}\label{f}
f=-\frac{T}{2\piup}\int_{-\piup}^{\piup}\log\left[2\cosh\left(\frac{E_{\gamma}(k)}{2T}\right)\right]\rd k,
\end{eqnarray}
here, $T$ is the temperature, and the Boltzmann constant was set to unity ($k_{\text B} = 1$) for the sake of simplicity. Using standard relations, one can easily obtain expressions for magnetization and polarization of the system:
\begin{eqnarray}\label{mp}
m=\frac{1}{4\piup}\int_{-\piup}^{\piup}\tanh\left(\frac{E_{\gamma}(k)}{2T}\right)\frac{B-J\cos k}{\lambda_k}\rd k,\;\;\;
p=\frac{1}{4\piup}\int_{-\piup}^{\piup}\tanh\left(\frac{E_{\gamma}(k)}{2T}\right)\sin k \rd k. 
\end{eqnarray}
Furthermore, the mixed magnetoelectric susceptibility is useful for figuring out important properties of the MEE:
\begin{eqnarray}\label{al}
\alpha=\frac{1}{8\piup T}\int_{-\piup}^{\piup}\frac{\left(B-J\cos k\right)\sin k}{\lambda_k\cosh^2\left(\frac{E_{\gamma}(k)}{2T}\right)}\rd k.
\end{eqnarray}
Particularly, we are going to figure out an effect of $XY$ anisotropy parameter $\gamma$ on the MEE. In case of a vanishing electric field, the spectrum of the model is always non-negative, thus, the system is always empty (in terms of Bogoliubov quasi-particles), and an increasing $XY$ anisotropy  always decreases the magnetization. Polarization in this case is zero. In virtue of DM terms and electric field, the spin-1/2 $XY$ chain with KNB mechanism features non-monotonous behavior of magnetization as a function of $\gamma$. In the figure \ref{fig3}, the polarization and magnetization dependence on $XY$ anisotropy $\gamma$ are exhibited. As in the case of finite $\gamma$, the system can have a gapless spectrum as well as a gapped one depending on the mutual relation between the electric field, magnetic field and $XY$ anisotropy. Therefore, the behavior of local observables is also non-monotonous. In contrast to the $E=0$ case, one can see magnetization growing with $\gamma$ [figure \ref{fig3} (right-hand panel)] within the gapless phase. Once the value of $XY$ anisotropy crosses the critical value,
\begin{eqnarray}\label{gc}
|\gamma_c|=\frac{1}{|J|}\sqrt{J^2+E^2-B^2},
\end{eqnarray}
a gap opens and magnetization starts to decrease (figure \ref{fig3}, right-hand panel, blue dashed and black dot-dashed lines). If the value of the magnetic field is greater than $\sqrt{E^2+J^2}$ (figure \ref{fig3}, right-hand panel, magenta dotted line), there is no gapless phase and magnetization exhibits a monotonous decrease with increasing $\gamma$.
Behavior of  polarization as a function of $\gamma$ (figure \ref{fig3}, left-hand panel) shares much in common with the effect of the magnetic field, a monotonous decrease in the gapless phase with plateau at zero for a gapped phase. Red solid and blue dashed lines go to zero at the same value of $\gamma$, as for $B^2 \leqslant J^2$, the value of $\gamma$ at which the free-fermion states start to fill up is the same, $J^2\gamma^2=E^2$.  In figure \ref{fig4} polarization (left-hand panel) and magnetization (right-hand panel) dependence on the magnetic field are illustrated. Though the behaviour of magnetization of the $XY$ chain is well known and understandable, here an additional feature can be pointed out.
 For $E=0$ and non-zero $\gamma$, besides the absence of the saturation field there is only one phase without any features on the magnetization curve. In case of finite $E$, it is possible to have both smooth magnetization curve for $E^2<J^2\gamma^2$ (figure \ref{fig4} right-hand panel, red solid line) as well as the curve with a cusp corresponding to transition from gapless regime to the gapped one (figure \ref{fig4} right-hand panel, blue dashed, black dot-dashed and magenta dotted lines). Interestingly, the magnetization curves are exactly the same for all values of $E^2<J^2\gamma^2$, since in all these cases the spectrum touches zero at one single point. For the values of electric field $E^2\geqslant J^2\gamma^2$, the magnetization curves have a cusp at $B_c=\sqrt{E^2+J^2(1-\gamma^2)}$ separating the gapless regime from the gapped one (figure \ref{fig4} right-hand panel, blue dashed, black dot-dashed and magenta dotted lines). The left-hand panel of the figure \ref{fig4} demonstrates the magnetic-field effect on  polarization. Four curves are presented for four different constant values of the electric field. All three curves share the same qualitative trend, a monotonous decrease from maximal values at $B=0$ to zero at $B_c=\sqrt{E^2+J^2(1-\gamma^2)}$.

\begin{figure}[!t]
\centerline{\includegraphics[width=0.95\textwidth]{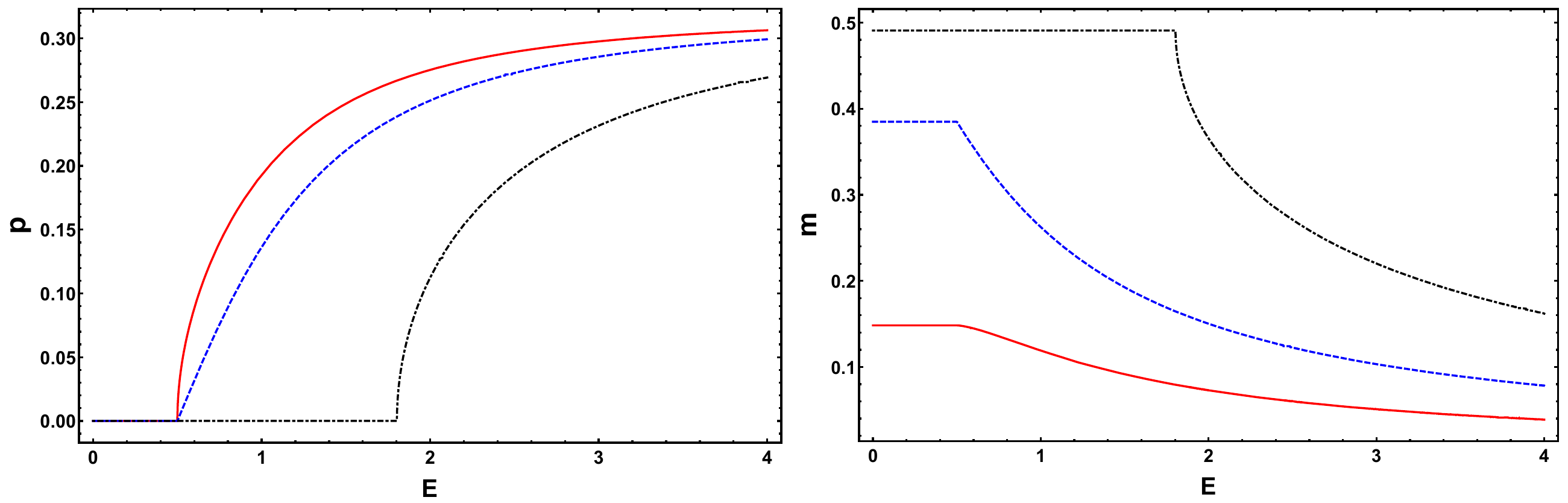}}
\caption{(Colour online) Low-temperature polarization (left-hand panel) and magnetization (right-hand panel) dependance on the electric field for the spin-1/2 XY
chain with KNB mechanism for $J = 1$, $T=0.0001$ and $\gamma = 0.5$. For both panels, $B=0.5$ (red solid line), $B=1$ (blue dashed line) and $B=2$ (black dot-dashed line).} \label{fig5}
\end{figure}

\begin{figure}[htb]
\centerline{\includegraphics[width=0.95\textwidth]{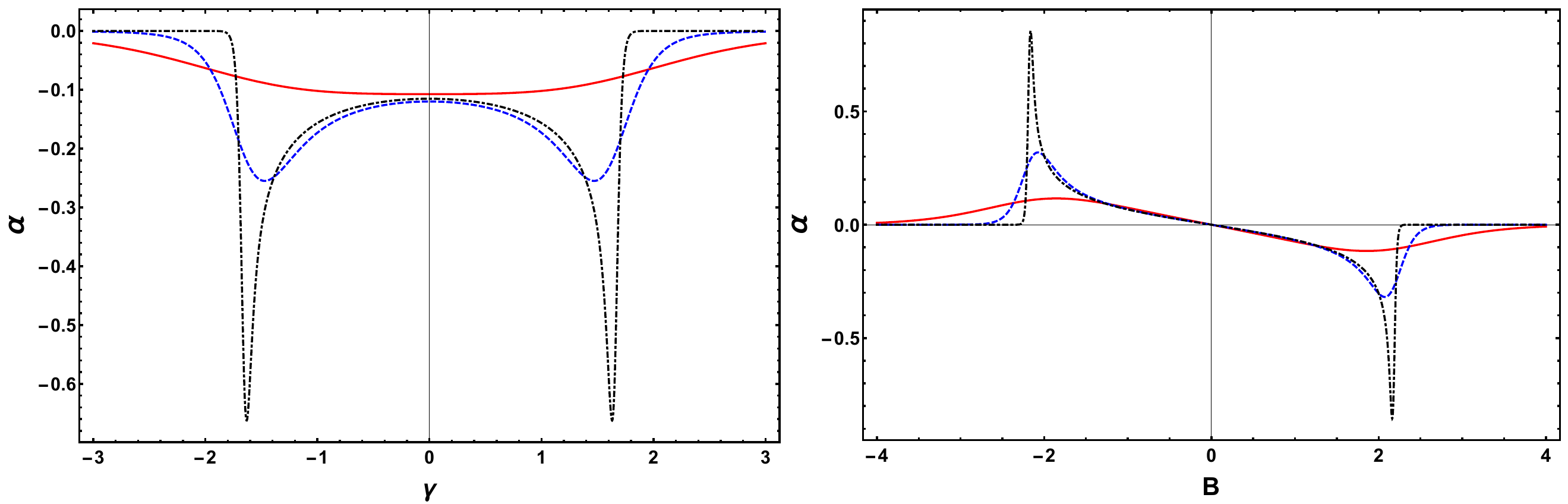}}
\caption{(Colour online) Magnetoelectric susceptibility dependance on $\gamma$ (left-hand panel) and magnetic field (right-hand panel) for $J=1$ and $E=2$ in appropriate units. For the left-hand panel $B=1.5$, $T=0.5$ (red solid line), $T=0.1$ (blue dashed line) and $T=0.015$ (black dot-dashed line). For the right-hand panel $\gamma=0.5$, $T=0.5$ (red solid line), $T=0.1$ (blue dashed line) and $T=0.015$ (black dot-dashed line).} \label{fig6}
\end{figure}

 The electric-field dependence of polarization and magnetization is presented in  figure~\ref{fig5}. Here again, one can distinguish two parts of the curve, corresponding to gapless and gapped regimes, respectively. The transition takes place at $E_c=\sqrt{B^2-J^2(1-\gamma^2)}$. The appearance of the critical point causes a thermal singularity in the behavior of susceptibilities. Considering magnetoelectric susceptibility given by equations (\ref{al_def}) and (\ref{al}), one can see well pronounced peaks at the corresponding values of $\gamma$ given by equation (\ref{gc}) (see figure~\ref{fig6}). Left-hand panel shows the $\gamma$ dependence of the magnetoelectric susceptibility for $E=2$, $B=1.5$ and three different temperatures, $T=0.5$, $0.1$ and $0.015$. Since the magnetization and polarization are strongly competing, $\alpha$ is always negative (for positive fields).  The development of peaks (negative) corresponding to the critical value of $\gamma$ is well pronounced here. The peaks are gradually smearing out with an increasing temperature. Thus, one can see that within the gapless phase the absolute value of the magnetoelectric susceptibility  grows with increasing $\gamma$ reaching a peak at the transition from the gapless regime to the gapped one. The peak shows a tendency to reach a diverging singularity at $T\rightarrow 0$. The same pattern can be seen in the magnetic field dependence (right-hand panel), according to the aforementioned property, the peak for the negative value of the magnetic field is positive.

  Interestingly, an important particular value of the $XY$ anisotropy, corresponding to the so-called quantum Ising chain \cite{pfu70}, does not have any specific feature in the sense of MEE. All the results concerning MEE are qualitatively similar to the ones presented for the spin-1/2 XY chain with $\gamma \neq 0$. The quantitative difference consists in the form of parameters region which corresponds to the gapless regime (or non-zero polarization region). In the case of $\gamma=1$, the region is given by the following conditions: $E^2 \geqslant J^2$ and $B^2 \leqslant E^2$.

\section{Conclusion}
In the present paper we considered MEE in the exactly solvable spin-1/2 $XY$ chain with KNB mechanism. Our main goal was to figure out the interplay between $XY$ anisotropy $\gamma$ and MEE. It turned out that the main difference from the properties of underlying $XY$ chain stems out from the fact that the appearance of an effective DM term in the Hamiltonian makes the gapless structure of the spectrum possible in some region of the parameter
space, which is determined by electric field, magnetic field and $XY$ anisotropy. Thus, even for the ordinary magnetization curve, the interplay between $XY$ anisotropy and electric field (effective DM term) brings essential modifications. It is well known that in case of $E=0$, the magnetization curve is always smooth, and there are no saturation phenomena. For the spin-1/2 $XY$ chain with KNB mechanism, this is still the case for
weak electric fields, but the situation drastically changes  for $E^2 > J^2\gamma^2$ when a transition point (cusp) appears in the magnetization curve.
This critical point corresponds to the transition from a gapless to a gapped spectrum and emerges under the conditions $B^2=E^2+J^2(1-\gamma^2)$ and $E^2>J^2\gamma^2$. Furthermore, the influence of the $XY$-anisotropy on the behaviour of the magnetization curve is essentially different for the gapless and the gapped phases. As far as the spectrum is gapless, magnetization grows with increasing $\gamma$. For the gapped phase, $XY$ anisotropy makes an opposite contribution to magnetization. Since polarization is always zero for a gapped situation in our model, $\gamma$ can affect polarization only within the gapless phase, where it  decreases with increasing $\gamma$.
Magnetoelectric susceptibility is shown to have a characteristic peak at a critical point closely associated with the opening of a spin gap in
the excitation spectrum. We also presented a zero-temperature description of the MEE for the spin-1/2 $XX$ chain being a particular case without XY
anisotropy ($\gamma = 0$), which exhibits the simplest possible MEE in quantum spin chains.

\section*{Acknowledgements}
The author expresses his deep gratitude to Taras Verkholyak and Artem Badasyan for helpful discussions.

\ukrainianpart

\title{Повна назва: Зразок статті та поради авторам}
\author{А.В. Тор\refaddr{label1,label2}, Б.В. Тор\refaddr{label2}}
\addresses{
\addr{label1} Університет ім. Орнштейна, Софтленд, 10041 Цельсій, вул. Реін, 1
\addr{label2} Інститут ім. Церніке, Солідшир, 20451 Фаренгейт, пр. Рівер, 2
}
%
%
%
\newpage
\ukrainianpart

\title{Вплив $XY$ анізотропії на магнетоелектричний ефект у спін-1/2 $XY$ ланцюжку в поперечному полі}
\author{В. Оганян}
\address{
 Лабораторія теоретичної фізики і Об'єднана лабораторія теоретичної фізики --- Філіал Міжнародного центру теоретичної фізики ім. Абдуса Салама у Вірменії, Єреванський державний університет,\\ вул. Манукяна 1, Єреван 0025, Вірменія
}

\makeukrtitle

\begin{abstract}
Розглянуто магнетоелектричний ефект у спін-1/2 $XY$ ланцюжку у поперечному полі, який виникає відповідно до механізму Кацури-Наґаоси-Балацького. Просторову орієнтацію електричного поля вибрано таким чином, щоб забезпечити точний розв'язок моделі у термінах вільних безспінових ферміонів. Розглядається найпростіша модель квантового спінового ланцюжка, яка демонструє магнетоелектричний ефект через механізм Кацури-Наґаоси-Балацького. Модель демонструє якомога простішу форму намагніченості, поляризації і сприйнятливості в залежності від електричного та магнітного полів. Для випадку довільної $XY$ анізотропії виявлено немонотонну залежність намагніченості від параметра $XY$ анізотропії. Ця неоднорідна поведінка зумовлена критичною точкою, яка пов'язана з можливістю контролювати появу щілинних або безщілинних станів за допомогою електричного поля. При критичних значеннях параметрів системи продемонстровано сингулярності магнетоелектричної сприйнятливості.

%
\keywords механізм Кацури-Наґаоси-Балацького, магнетоелектричний ефект, $XY$ ланцюжок, вільні безспінові ферміони
\end{abstract}

\end{document}